\documentclass[seceq]{ptptex}

\usepackage{graphicx}

\newcommand{\sph}	{{\rm sph}}
\newcommand{\weak}	{{\rm W}}

\newcommand{\scalar}	{{\rm S}}
\newcommand{\elemag}	{{\rm EM}}
\newcommand{\ELE}	{{\rm E}}
\newcommand{\MAG}	{{\rm M}}

\newcommand{\Hamil}	{{\cal H}}
\newcommand{\fs}	{{\rm 1s}}

\newcommand{\sing}	{{\rm sing}}

\newcommand{\diag}	{{\rm diag}}
\newcommand{\BG}	{{\rm BG}}
\newcommand{\vac}	{{\rm vac}}

\newcommand{\BH}	{{\rm BH}}
\newcommand{\pl}	{{\rm pl}}

\newcommand{\GUT}	{{\rm GUT}}

\newcommand{\gnear}	{{{\begin{array}{c} > \\[-0.8em] \vspace{-0.25em}\sim \end{array}}}}
\newcommand{\GeV}	{{\rm \;GeV}}

\newcommand{\fig}[1]	{Fig.~\ref{#1}}
\newcommand{\EQ}[1]	{(\ref{#1})}

\title{
Atomic Structure in Black Hole%
}

\author{
Yukinori \textsc{Nagatani}%
\footnote{E-mail: yukinori\_nagatani (at) pref.okayama.jp.
This article is based on a talk given by the present author
at the international workshop
``Frontiers of Quantum Physics'', February 17-19, 2005,
Yukawa Institute for Theoretical Physics, Kyoto University, Japan.
}%
}

\inst{
Okayama Institute for Quantum Physics,\\
Kyoyama 1-9-1, Okayama 700-0015, Japan
}

\abst{
We propose that any black hole has atomic structure in its inside
and has no horizon as a model of black holes.
Our proposal is founded on a mean field approximation of gravity.
The structure of our model consists of
a (charged) singularity at the center and
quantum fluctuations of fields around the singularity,
namely, it is quite similar to that of atoms.
Any properties of black holes, e.g. entropy, can be explained by the model.
The model naturally quantizes black holes.
In particular,
we find the minimum black hole,
whose structure is similar to that of the hydrogen atom 
and whose Schwarzschild radius is
approximately $1.1287$ times the Planck length.
Our approach is conceptually similar to the Bohr's model
of the atomic structure,
and the concept of the minimum Schwarzschild radius is similar to
that of the Bohr radius.
The model predicts that black holes carry baryon number,
and the baryon number is rapidly violated.
This baryon number violation can be used as verification of the model.
}

\begin{document}

\maketitle

\section{Introduction}

The most natural way to solve the information paradox for black holes
is to construct a {\em structure model} of black holes.
If we construct a model of a spherical object
whose properties,
e.g., mass, charges, radius, entropy \cite{Bekenstein:1973ur},
emitted radiation \cite{Hawking:1975sw:Hawking:1974rv}, and so on,
correspond to those of a black hole,
the object can be identified as the black hole for distant observers.
If the object has no horizon and has an interior structure,
there arises no information paradox
\cite{Hotta:1997yj,Iizuka:2003ad,Nagatani:2003rj,Nagatani:2003new,Nagatani:2003new2,Nagatani:2005qe}.
Such an object is the structure model of a black hole.
K.~Hotta proposed the Planck solid ball model \cite{Hotta:1997yj},
which is identical to the structure model of black holes.
According to the Planck solid ball model,
any black hole consists of a ball consisting of the Planck solid and
a layer of thermal radiation around the ball.
The Planck solid is hypothetical matter
which arises through the stringy thermal phase transition
due to the high temperature of the radiation.
The temperature of the radiation around the ball
becomes very high,
because the deep gravitational potential [small $g_{00}(r)$]
creates a blue-shift effect [$T(r) \propto 1/\sqrt{g_{00}(r)}$]
\cite{Hotta:1997yj,Iizuka:2003ad,Nagatani:2003ps}.
The entropy of the black holes is carried by the radiation.
The Hawking radiation is explained as a leak of the thermal radiation.

\begin{figure}
 \begin{center}
  \begin{tabular}{@{}l@{\hspace{2mm}}c@{}}
   \includegraphics[width=65mm]{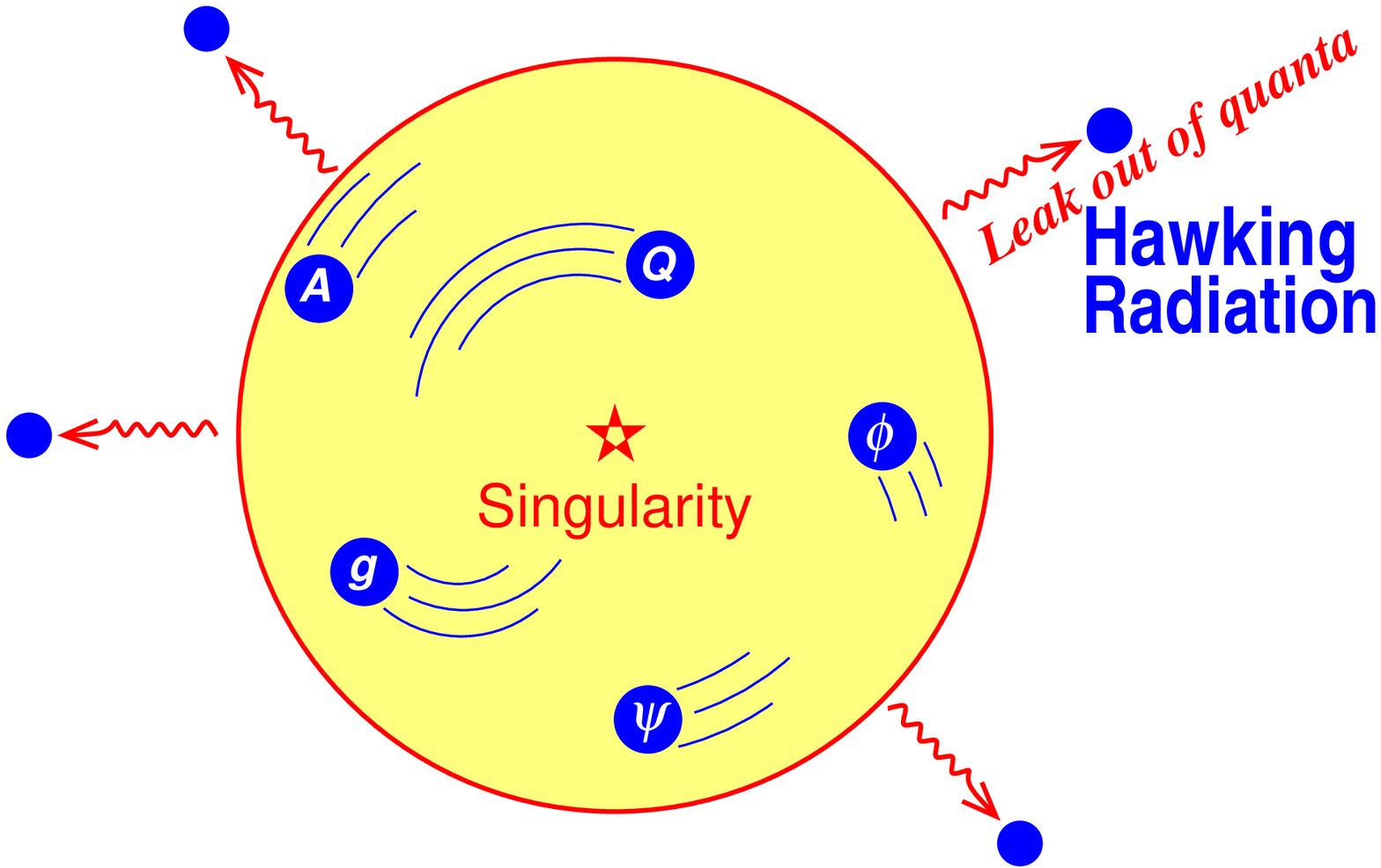}  &
   \includegraphics[width=65mm]{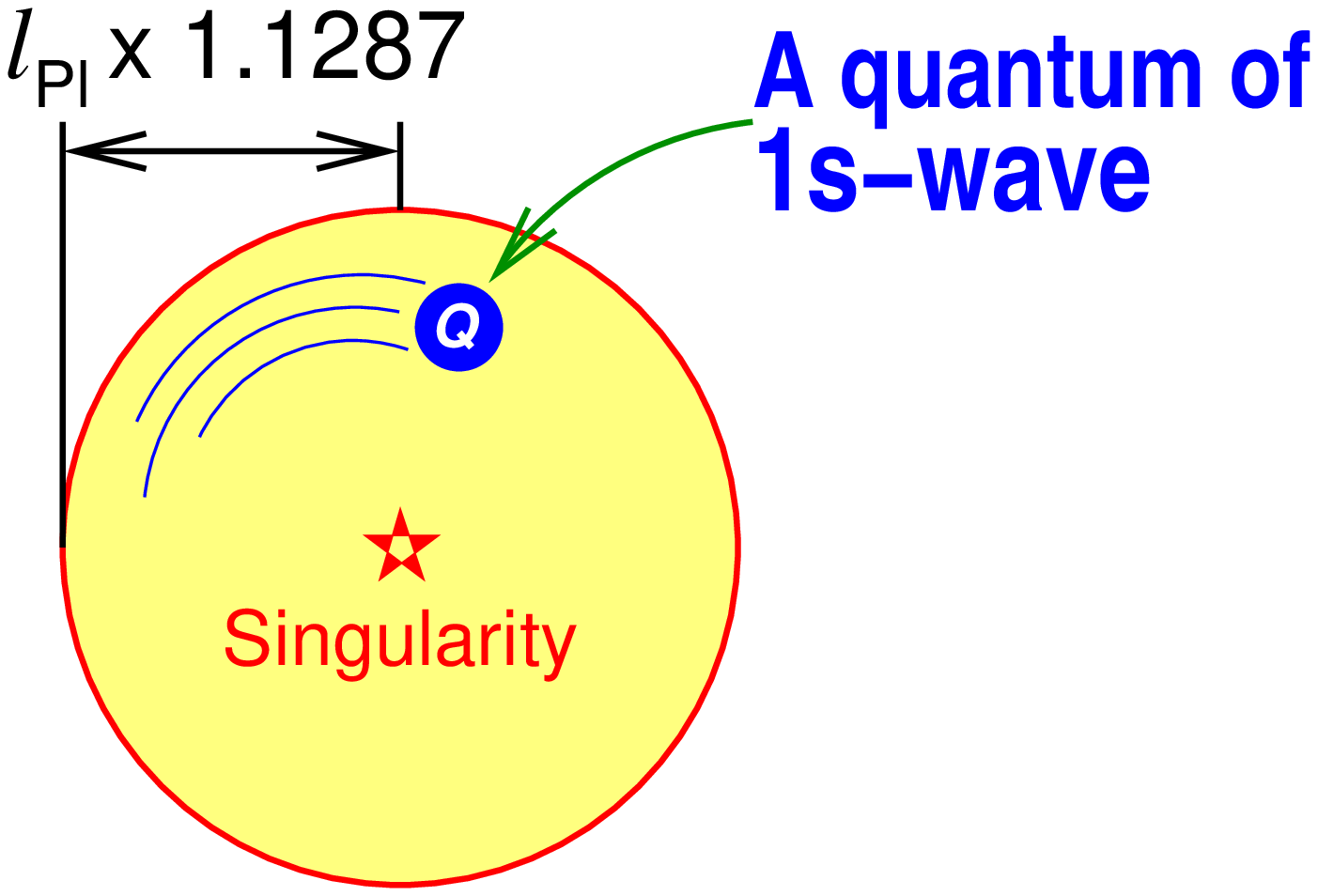}\\
   a) \parbox[t]{50mm}{The atomic structure of a generic black hole.} &
   b) \parbox[t]{50mm}{The hydrogen atomic structure of
   the minimum black hole.}
  \end{tabular}
 \end{center}
 \caption{%
 The structure model of black holes.
 \label{Fig.Atom.eps}%
 }%
\end{figure}

Based on the Planck solid ball model
without the Planck solid transition hypothesis,
we have proposed
the radiation ball model
\cite{Nagatani:2003rj,Nagatani:2003new,Nagatani:2003new2}
and the atomic structure model
\cite{Nagatani:2005qe}.
According to these structure model,
any black hole consists of a central singularity and
quantum fluctuations around the singularity.
The generic structure of the model is shown in \fig{Fig.Atom.eps}-a.
In this talk, we present our concept of the structure model.
The results and prediction of the models are the following:
Black holes are naturally quantized,
and especially the minimum black hole is found out.
The structure of the minimum black hole is similar
to that of a hydrogen atom (see \fig{Fig.Atom.eps}-b),
and its radius, namely the minimum Schwarzschild radius,
is approximately $1.1287$ times the Planck length.
The entropy of the structure is carried by the quantum fluctuations.
When the radius of the structure is much larger than Planck length,
the entropy of the structure
is proportional to the surface area of the structure and
is almost the same as the Bekenstein entropy.
Black holes can carry baryon number,
and the Hawking radiation contains the baryon number.
The baryon number is rapidly decaying
due to the baryon violating interactions in the structure.
Our model can be verified 
by detecting the baryon radiation and its decay curve.

\section{Concept of The Structure Model of Black Holes}

Our concept of the structure model of black holes
is the following \cite{Nagatani:2005qe}:
We assume Einstein gravity, an electromagnetic field
and several other fields.
The black hole consists of both a singularity at the center
and fluctuating fields around the singularity.
The singularity possesses electric (magnetic) charge and
creates a spherically symmetric time independent electric (magnetic) field
around the singularity.
The charge of the singularity
causes the singularity to be gravitationally-repulsive,
as described by the Einstein equation.
The fluctuating fields contain all species of fields in the theory.
The repulsive singularity, the static electromagnetic field and
the fluctuating fields create
a gravitational field as a mean field
which obeys the Einstein equation.
The gravitational field
binds the field fluctuation into a ball structure.
The radius of the structure is identical to the Schwarzschild radius
of the corresponding charged black hole.
The exterior of the structure corresponds
to that of the ordinary black hole.
The form of the field fluctuations in the structure
is determined by
quantizing the fluctuating fields
in the mean field of the gravity background.
The right-hand side of the Einstein equation
is the expectation value
of the energy-momentum tensor for the quantum-fluctuating fields.

\section{Target Black Hole of the Model Construction}

We construct the structure model
of the Reissner-Nordstr\"om (RN) black hole
whose Schwarzschild radius is $r_\BH$, and whose charge is $Q$.
We introduce a charge-parameter $q := Q/(m_\pl r_\BH)$
which has a value from 0 to 1,
and $q \rightarrow 1$ indicates the extremal limit.
The metric of the RN black hole is given by
\begin{eqnarray}
 ds^2
  &=&
  F_\BH(r) dt^2
  - G_\BH(r) dr^2
  - r^2 d\theta^2
  - r^2 \sin^2\theta \, d\varphi^2,
  \label{RN-metric.eq}
\end{eqnarray}
where
\begin{eqnarray}
 F_\BH(r) &=&
  \left(1 - \frac{r_\BH}{r}\right)
  \left(1 - q^2 \frac{r_\BH}{r}\right),
  \qquad
 G_\BH(r) = \frac{1}{F_\BH(r)},
  \label{RN.eq}
\end{eqnarray}
are metric elements.
The metric of the exterior region $(r \gnear r_\BH)$
of the structure model
should agree with that of the black hole metric (\ref{RN-metric.eq}).
The charge of the structure model also should correspond
with that of the black hole.

The Hawking temperature of the charged black hole becomes
\begin{eqnarray}
  T_\BH &=& \frac{1}{4\pi} \frac{1}{r_\BH} (1 - q^2).
  \label{Hawking-temp.eq}
\end{eqnarray}
When $q=1$, the black hole becomes the extremal
and its Hawking temperature becomes zero.
We put the black hole into the background of the radiation
with the temperature $T_\BH$
to consider the stationary situation and the equilibrium of the system
\cite{Gibbons:1976ue}.
The energy density of the background radiation ($r\rightarrow\infty$)
is given by the thermodynamical relation:
\begin{eqnarray}
 \rho_\BG &=& \frac{\pi^2}{30} g_* T_\BH^4,
  \label{rho-BG.eq}
\end{eqnarray}
where $g_*$ is the degree of the freedom of the radiation.
We assume that $g_*$ is a constant for simplicity.

Here we should introduce the positive cosmological constant
$\Lambda = (8\pi/m_\pl^2) \rho_\vac$
to stabilize the background universe from the effect of 
constant energy density $\rho_\BG$.
The background space-time becomes the Einstein static universe 
by choosing $\rho_\vac = \rho_\BG$.

\section{Mean Field Approximation of Gravity}

One of the most important concept in the structure model
is {\em the mean field approximation of gravity}.
The approximation is defined by decomposition of fields
in the base theory which we consider.
When there are
a gravity field (metric) $g_{\mu\nu}$,
an electromagnetic field $A_\mu$,
a scalar field $\phi$,
matter fields $\psi_i$,
and so on in the theory,
any field in the theory is decomposed
into a spherically-symmetric time-independent part
and a fluctuating part:
\begin{eqnarray}
 &&
 \begin{array}[c]{ccccc}
         g_{\mu\nu}(t,r,\theta,\varphi) &=& g_{\mu\nu}(r) &+&
  \delta g_{\mu\nu}(t,r,\theta,\varphi) \\
         A_{\mu}(t,r,\theta,\varphi)    &=& A_{\mu}(r)    &+&
  \delta A_{\mu}(t,r,\theta,\varphi)    \\
         \phi(t,r,\theta,\varphi)       &=& 0             &+&
  \delta \phi(t,r,\theta,\varphi)       \\
         \psi_i(t,r,\theta,\varphi)     &=& 0             &+&
  \delta \psi_i(t,r,\theta,\varphi)     \\
  \vdots                                & & \vdots        & &
  \vdots                                \\
  & & \mbox{\small mean field} & & \mbox{\small fluctuations}
 \end{array},
 \label{decomposition.eq}
\end{eqnarray}
where $r,\theta$ and $\varphi$ are polar coordinates.
The time independent parts are simply mean fields.
The metric $g_{\mu\nu}(r)$ describes a background geometry
as a mean field.
All fields exist in the background space-time
described by the mean field metric $g_{\mu\nu}(r)$.
The electromagnetic field $A_{\mu}(r)$ also describes a background
field.
The metric fluctuation $\delta g_{\mu\nu}$ describes gravitons
in the background.
The other fluctuations $\delta A_\mu$, $\delta\phi$, $\delta\psi$,
$\cdots$
correspond with photons, scalar bosons, matter particles and so on.
All of these fluctuations
$\delta g_{\mu\nu}$, $\delta A_\mu$, $\delta\phi$, $\delta\psi$, $\cdots$
are represented by $\Phi_i$.

In our mean field approximation of gravity,
the system is described by combination of the mean field
$(g_{\mu\nu}(r),\; A_\mu(r))$ and the fluctuation fields $\Phi_i$.
The fluctuation $\Phi_i$ is quantized in the background geometry
$g_{\mu\nu}(r)$, and
the mean field gravity $g_{\mu\nu}(r)$ is determined
by the energy-momentum tensor of the fluctuation $\Phi_i$.
The essence of our approach is that
the balance amang the mean field of gravity and the fluctuations
determines the structure of the model of black holes.
When we construct a model of the black hole,
we require that the exterior region $(r \gnear r_\BH)$ of the structure
corresponds with that of the ordinary black hole as a boundary condition.
Thus,
for the exterior region $(r \gnear r_\BH)$,
the metric $g_{\mu\nu}(r)$ should correspond
with the metric of the corresponding black hole.

The general form of the metric $g_{\mu\nu}(r)$ of the mean field
becomes
\begin{eqnarray}
  ds^2 &=&
  F(r) dt^2 \;-\; G(r) dr^2
  \;-\; r^2 d\theta^2 \;-\; r^2\sin^2\theta d\varphi^2,
  \label{metric.eq}
\end{eqnarray}
due to the spherical symmetry and time independence.
The space time structure of the model
is determined by the metric elements $F(r)$ and $G(r)$,
and these metric elements are determined by the Einstein equation:
\begin{eqnarray}
  R_{\mu\nu}
   - \frac{1}{2} g_{\mu\nu} R
   - \Lambda\delta_{\mu\nu}
  &=& \kappa
  \left\{
   T_{\elemag\mu\nu}
   +
   \sum_{i}
   \left<\Phi| T_{\mu\nu}[\Phi_i] |\Phi\right>
  \right\}.
  \label{ein0.eq}
\end{eqnarray}
The left-hand side of (\ref{ein0.eq})
is the Einstein tensor of the metric (\ref{metric.eq}).
The first term of the right-hand side of (\ref{ein0.eq})
is the energy-momentum tensor
of the background electromagnetic fields $A_\mu(r)$.
The second term is sum of the energy-momentum tensors
of the all fluctuations.
Because we construct a model of non-rotating stationary black hole,
we only consider the quantum state of fluctuations $|\Phi\rangle$
such that the sum of the energy-momentum tensors
$\sum_{i}\left<\Phi| T_{\mu}^{\ \nu}[\Phi_i] |\Phi\right>$
becomes diagonal and time-independent.

We assume that the charge density of the total fluctuations
is neutral for simplicity.
Under this assumption of the neutrality of the fluctuations,
the Maxwell equation for the electromagnetic field $A_{\mu}(r)$ is solved,
and the solution of the field strength $F_{\mu\nu}$ is given by
\begin{eqnarray}
  F_{\mu\nu} dx^\mu \wedge dx^\nu
  &=& E(r) \sqrt{FG} dt \wedge dr
  \;+\; B(r) r^2 \sin\theta d\theta \wedge d\varphi,\nonumber
\end{eqnarray}
where 
\begin{eqnarray}
  E(r) \;:=\; \frac{Q_\ELE}{r^2}, \qquad
  B(r) \;:=\; \frac{Q_\MAG}{r^2}
  \label{EM-Field.eq}
\end{eqnarray}
are the radial elements of the electric
and magnetic fields, respectively.
The constants $Q_\ELE$ and $Q_\MAG$ are the electric and magnetic
charges respectively at the central singularity $(r=0)$.
The non-trivial elements of the energy-momentum tensor for the
electromagnetic field \EQ{EM-Field.eq} are
\begin{eqnarray}
  T_{\elemag 0}^{\ \ \ \ 0} \ =\ 
  T_{\elemag r}^{\ \ \ \ r} \ =\ 
 -T_{\elemag \theta}^{\ \ \ \ \theta} \ =\ 
 -T_{\elemag \varphi}^{\ \ \ \ \varphi} \ =\ 
 \frac{Q^2}{8\pi}\frac{1}{r^4},
\end{eqnarray}
where we have defined $Q^2 := Q_\ELE^2 + Q_\MAG^2$.

\section{Quantization of Fluctuations}

We regard the fluctuations $\Phi_i$ as free fields
in the gravity background,
because
the gravitational interaction to the fields is essential
and the other interactions among the fields are not important
in our approach.
Therefore, 
the fluctuation fields $\Phi_i$ simply obey
the free wave equations in the background gravity $g_{\mu\nu}(r)$,
as an essential approximation of the model construction.

The bosonic fields of the fluctuations obey
\begin{eqnarray}
 \Box_{(g_{\mu\nu})} \Phi_i &=& 0,
  \label{wave-equation.eq}
\end{eqnarray}
where we have defined the d'Alembertian
$
\Box_{(g_{\mu\nu})} f :=
\frac{1}{\sqrt{g}} \partial_\mu
\left[ \sqrt{g} g^{\mu\nu} \partial_\nu f \right]
$
in the background metric $g_{\mu\nu}$.
The energy-momentum tensor of the field $\Phi_i$ becomes
\begin{eqnarray}
 T_{\mu\nu}[\Phi_i]
  &=&
  \frac{1}{2}
  \left(
   \partial_\mu\Phi^\dagger_i \partial_\nu\Phi_i
   +
   \partial_\nu\Phi^\dagger_i \partial_\mu\Phi_i
  \right)
   - \frac{1}{2} g_{\mu\nu} g^{\alpha\beta}
   \partial_\alpha\Phi^\dagger_i \partial_\beta\Phi_i.
   \label{energy-momentum-tensor.eq}
\end{eqnarray}

Quantization of the fluctuation fields are done by
the canonical quantization formalism.
The Hamiltonian operator is given by
\begin{eqnarray}
 \Hamil &:=& \sum_i \int dr d\theta d\varphi \sqrt{FG} r^2 \sin\theta \ 
  T_{0}^{\ 0}[\Phi_i],
  \label{Hamiltonian-1.eq}
\end{eqnarray}
and the conjugate momentum of the field $\Phi_i(t,r,\theta,\varphi)$
is given by
\begin{eqnarray}
 \Pi_i
  &:=&
  \sqrt{\frac{G}{F}} r^2 \sin\theta \; \partial_0{\Phi_i}.
\end{eqnarray}
By employing the equal-time commutation relations:
\begin{eqnarray}
 \left[
  \Phi_i(t,r_1,\theta_1,\varphi_1),
  \Pi_j(t,r_2,\theta_2,\varphi_2)
 \right] &=&  i
 \delta_{ij}
 \delta(r_1 - r_2) \delta(\theta_1 -\theta_2) \delta(\varphi_1 - \varphi_2),
 \nonumber\\
 \left[
  \Phi_i(t,r_1,\theta_1,\varphi_1),
  \Phi_j(t,r_2,\theta_2,\varphi_2)
 \right] &=&  0,
 \nonumber\\
 \left[
  \Pi_i(t,r_1,\theta_1,\varphi_1),
  \Pi_j(t,r_2,\theta_2,\varphi_2)
 \right] &=&  0,
\end{eqnarray}%
the fluctuation fields are quantized.

By using the real spherically harmonic function
$Y_l^{\ m}(\theta,\varphi)$
and the radial mode function $f_{nl}(r)$,
the mode expansion for the field operators is naturally given by
\begin{eqnarray}
 \Phi_i &=&
  \sum_{n=1}^{\infty} \sum_{l=0}^{\infty} \sum_{m=-l}^{l}
  \frac{1}{\sqrt{2 \omega_{nl}}}\;
  \left\{
  a_{inlm} \, e^{-i\omega_{nl} t} 
  \:+\:
  a_{inlm}^\dagger \, e^{+i\omega_{nl} t} 
  \right\}
  f_{nl}(r) \: Y_l^{\ m}(\theta,\varphi),\ \ 
  \label{mode-expansion.eq}
\end{eqnarray}
where the summation is performed over all of the physical modes $\{nlm\}$
in the background of the gravity \EQ{metric.eq}.
The index $n$ is numbering of the energy levels for given $l$.
In order for the expansion (\ref{mode-expansion.eq}) to be solution of
the wave equation (\ref{wave-equation.eq}),
the radial function $f_{nl}(r)$ should satisfy
\begin{eqnarray}
 \sqrt{\frac{F}{G}} \frac{1}{r^2} \partial_r
  \left[ \sqrt{\frac{F}{G}} r^2 \partial_r f_{nl} \right]
  \;+\; \omega_{nl}^2 f_{nl}
  \;-\; l(l+1) \frac{F}{r^2} f_{nl} &=& 0.
  \label{scalar_eom2.eq}
\end{eqnarray}
We require the normalization condition for $f_{nl}(r)$ as
\begin{eqnarray}
  \int_0^\infty dr \sqrt{\textstyle \frac{G}{F}} r^2 f_{nl}^2
  &=& 1.
  \label{C.eq}
\end{eqnarray}
for the suitable normalization of the creation and annihilation operators.
The boundary condition for $r \rightarrow 0$ is
$f_{nl} \rightarrow 0$ or $\partial_r f_{nl}(r) \rightarrow 0$,
and that for $r \rightarrow \infty$ is $f_{nl}(r) \rightarrow 0$.

Here, we find the following well-known commutation relations:
\begin{eqnarray}
 \left[a_{i n_1l_1m_1}, a_{j n_2l_2m_2}^\dagger\right]
  &=& \delta_{ij} \delta_{n_1n_2} \delta_{l_1l_2} \delta_{m_1m_2},
  \nonumber\\
 \left[a_{i n_1l_1m_1}, a_{j n_2l_2m_2} \right] &=& 0, \qquad
 \left[a_{i n_1l_1m_1}^\dagger, a_{j n_2l_2m_2}^\dagger \right] \ =\ 0.
\end{eqnarray}
The vacuum of the system is defined as $a_{nlm} \left|0\right> = 0$
for all physical modes $\{nlm\}$.
The Hamiltonian \EQ{Hamiltonian-1.eq} becomes
\begin{eqnarray}
 \Hamil
 &=&
  \sum_{i}
  \sum_{nlm} \omega_{nl}
  \left(
   a_{inlm}^\dagger a_{inlm}
   \;+\; \frac{1}{2}
  \right).
  \label{Hamiltonian-2.eq}
\end{eqnarray}
We ignore the vacuum energy
in the energy-momentum tensor (\ref{energy-momentum-tensor.eq})
and in the Hamiltonian (\ref{Hamiltonian-2.eq}) for simplicity.
This might be justified by assumption of supersymmetry of the theory.

The general state is given by
\begin{eqnarray}
 \left| \Phi \right> = \prod_{\{nlm\}} (a_{nlm}^\dagger)^{N_{nlm}}
  \left|0\right>
\end{eqnarray}
which makes $\left< \Phi | T_{\mu}^{\ \nu} | \Phi \right>$
to be diagonal.
The solution of the model is labeled by possible $\{N_{nlm}\}$,
and the concrete solution is given by 
a set of $F(r)$, $G(r)$ and $\{ f_{nlm}(r) \}$
which satisfy the Einstein equation (\ref{ein0.eq})
and the wave equations (\ref{scalar_eom2.eq}).

To obtain a general solution,
we should solve a lot of simultaneous differential equations
(\ref{scalar_eom2.eq}) with the Einstein equation (\ref{ein0.eq})
with the boundary conditions,
and it seems to be technically difficult.
However we can obtain
the solutions which consist of a few quanta,
because we only need to solve a few simultaneous differential equations.

\section{Minimal Solution}

The minimal solution~\cite{Nagatani:2005qe} consists of a single quantum
in the lowest-energy in s-wave mode, i.e.,
$N_{100} = 1$ and the other $N_{nlm} = 0$.
We denote the state as $\left|1s\right>$.
The structure of the minimal solution
is just the same as that of the Hydrogen atom.

For simplicity,
we consider a minimal model of 
the near-extremal ($q \rightarrow 1$) Reissner-Nordstr\"om black hole,
and its Hawking temperature is almost zero.
Thus we do not need to consider the cosmological constant $\Lambda$,
because of the absence of the background radiation.

The expectation value of the energy-momentum tensor for the state
$\left|\fs\right>$
is
\begin{eqnarray}
 \left<\fs\right| {\textstyle T_0^{\ 0}} \left|\fs\right>
  =
 -\left<\fs\right| {\textstyle T_r^{\ r}} \left|\fs\right>
  &=&
  \frac{1}{8\pi} \frac{1}{\omega_{10} C_{10}}
  \left[
   \frac{\omega_{10}^2}{F} f_{10}^2 \;+\; \frac{1}{G} (\partial_r f_{10})^2
  \right],
  \nonumber\\
 -\left<\fs\right| {\textstyle T_\theta^{\ \theta}} \left|\fs\right>
  =
 -\left<\fs\right| {\textstyle T_\varphi^{\ \varphi}} \left|\fs\right>
  &=&
  \frac{1}{8\pi} \frac{1}{\omega_{10} C_{10}}
  \left[
   \frac{\omega_{10}^2}{F} f_{10}^2 \;-\; \frac{1}{G} (\partial_r f_{10})^2
  \right],\nonumber\\
  \left<\fs\right| (\mbox{non-diagonal-parts}) \left|\fs\right> &=& 0.
   \label{1s-EM-tensor.eq}
\end{eqnarray}
The Einstein equation (\ref{ein0.eq}) becomes
\begin{eqnarray}
 \frac{-G + G^2 + r G'}{r^2 G^2}
  &=&
  \frac{8\pi}{m_\pl^2}
  \left\{
   +
   \frac{Q^2}{8\pi}\frac{1}{r^4} \;+\;
   \left<\fs\right| T_0^{\ 0} \left| \fs \right>
  \right\},
   \label{ein1.eq}
   \\
 \frac{F - FG + r F'}{r^2 F G}
  &=&
  \frac{8\pi}{m_\pl^2}
  \left\{
   -
   \frac{Q^2}{8\pi}\frac{1}{r^4} \;-\;
   \left<\fs\right| T_r^{\ r} \left| \fs \right>
  \right\},
   \label{ein2.eq}
\end{eqnarray}
where the prime represents differentiation with respect $r$.
By subtracting \EQ{ein2.eq} from \EQ{ein1.eq}
with the property
$ \left<\fs\right| T_0^{\ 0} \left| \fs \right>
= - \left<\fs\right| T_r^{\ r} \left| \fs \right>$,
we find the relation
\begin{eqnarray}
 G &=&
  \biggl({\frac{r}{2} \frac{H'}{H} + 1}\biggr)
  \biggl({1 - \frac{Q^2}{m_\pl^2 r^2}}\biggr)^{-1},
  \label{defG.eq}
\end{eqnarray}
where we have defined the new parameter
\begin{eqnarray}
 H &:=& \frac{F}{G}.
  \label{defH.eq}
\end{eqnarray}
By substituting (\ref{defG.eq}) into (\ref{ein1.eq}),
the Einstein equation \EQ{ein1.eq} becomes
\begin{eqnarray}
   \frac{2 H H'' - H'^2 + \frac{4}{r} H H'}{2 H + r H'}
   \;-\; 4 q^2 \frac{r_\BH^2}{r^4} \frac{H}{1 - q^2 \frac{r_\BH^2}{r^2}}
   \;-\; \frac{2}{m_\pl^2}
   \left( \omega_{10}^2 {f}^2 + H {f}'^2 \right)
   &=& 0.
   \label{ein1-3.eq}
\end{eqnarray}
The wave equation (\ref{scalar_eom2.eq}) for the radial mode function $f$
for s-wave becomes
\begin{eqnarray}
 H {f}''
 \;+\; \left(\frac{2}{r} H 
 \:+\: \frac{1}{2} H' \right) {f}'
 \;+\; \omega_{10}^2 {f}
 &=& 0.
 \label{scalar_eom3.eq}
\end{eqnarray}

The simultaneous equations \EQ{ein1-3.eq} and \EQ{scalar_eom3.eq}
are numerically solved with the initial conditions
\begin{eqnarray}
 &&
 F(r) \ \rightarrow\  F_\sing(r),\quad
 G(r) \ \rightarrow\  G_\sing(r),\quad
 {f}(r) \ \rightarrow\  {f}_\sing,\quad
 {f}'(r) \ \rightarrow\  0
\end{eqnarray}
for $r\rightarrow0$,
where we have defined the Reissner-Nordstr\"om (RN) type singularity:
$F_\sing(r) := 
\alpha^2 \times \frac{(1 - q^2)^2}{12960 \sqrt{2\pi} m_\pl^4 r_\BH^4}
q^2 \left({r_\BH}/{r}\right)^2$ and
$G_\sing(r) := \frac{1}{q^2} \left({r}/{r_\BH}\right)^2$.
The constant $\alpha$ is red-shift factor of the singularity.
We should choose the radius of the solution, $r_\BH$,
the initial value of the mode function, ${f}_\sing$,
and the frequency of the mode, $\omega_{10}$,
so as to satisfy the exterior boundary conditions
\begin{eqnarray}
 &&
 F(r) \ \rightarrow\  F_\BH(r),\quad
 G(r) \ \rightarrow\  G_\BH(r),\quad
 {f}(r) \ \rightarrow\  0
\end{eqnarray}
for $r > r_\BH$ and
the normalization condition for ${f}$ in \EQ{C.eq}.
These boundary conditions yield a quantization of $r_\BH$
as 
${f}$ should not cross zero for $0<r<r_\BH$,
because the function ${f}$ describes the lowest s-wave mode
(the $\fs$-mode).

\begin{figure}
 \begin{center}
  \includegraphics[scale=0.4]{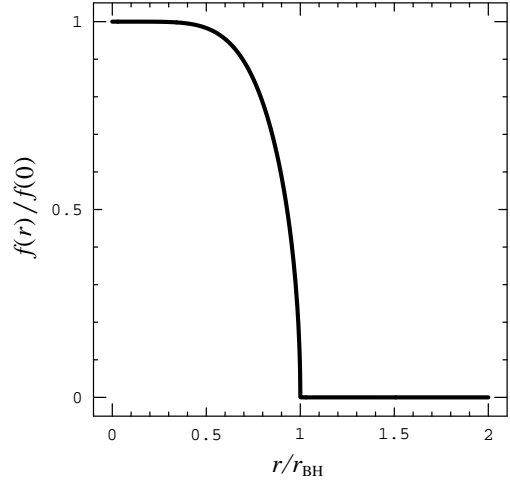}%
 \end{center}
 \caption{%
 Distribution of the radial mode function ${f}(r)$
 for the state $\left|\fs\right>$.
 This plot represents the quantum fluctuation of the field
 in the solution of the minimum structure model
 for a near-extremal black hole.
 While we chose the parameters as $q=0.99999$ and $\alpha=1$,
 the curve depends only weakly on the parameters.
 The vertical axis is normalized by ${f}(0) = 1.268776 \times m_\pl$.
 The horizontal axis represents the coordinate $r$ normalized by $r_\BH$.
 The radius of the solution,
 namely the minimal Schwarzschild radius,
 becomes $r_\BH = 1.12871 \times l_\pl$,
 where $l_\pl := 1/m_\pl$ is the Planck length.
 We find that the quantum fluctuation is completely bound into the structure.
 \label{Fig.f.eps}%
 }%
\end{figure}

\begin{figure}
 \begin{tabular}{@{}c@{\hspace{7.2mm}}c@{}}
  \includegraphics[scale=0.4]{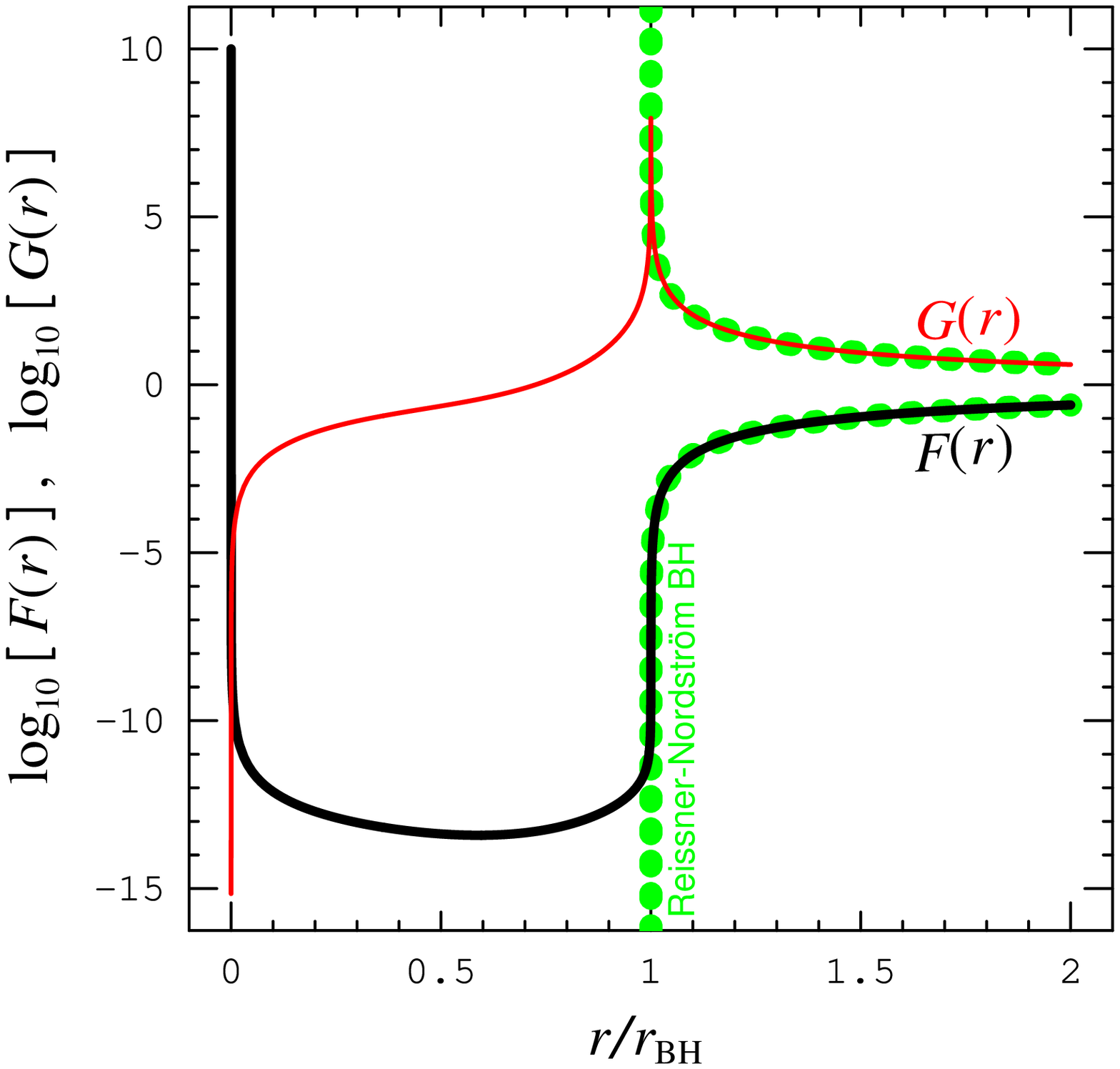}  &
  \includegraphics[scale=0.4]{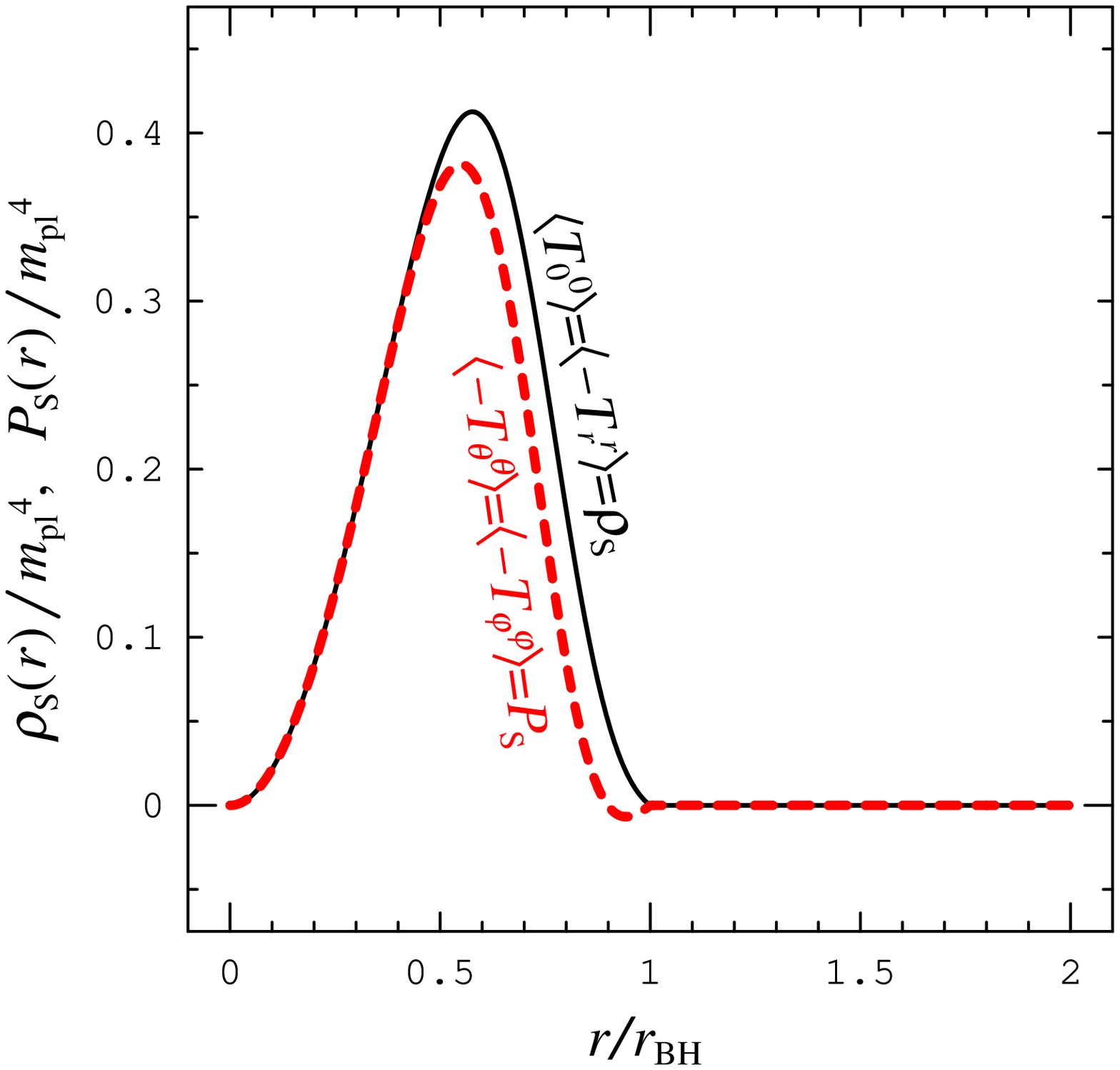}\\
  a) Elements of the metric $F(r)$ and $G(r)$ &
  b) Expectation values of 
 $\left<\fs\left| T_{\scalar\mu}^{\ \ \nu} \right|\fs\right>$
 \end{tabular}
 \caption{%
 a) The metric elements
 and b) the energy momentum tensor for the quantized field
 in the solution of the minimum model.
 a) 
 The thick solid curve is $F(r)$, and
 the thin solid curve is $G(r)$.
 The thick dotted curves indicate
 the exterior part ($r>r_\BH$) of the Reissner-Nordstr\"om metric
 [$F_\BH(r)$ and $G_\BH(r)$ in \EQ{RN.eq}]
 with the same radius $r_\BH$ and with the same charge $q$.
 The depth of the potential well [$F(r)$ for $0<r<r_\BH$]
 depends on $q$ and $\alpha$.
 b)
 The thin solid curve represents
 $\rho_\scalar
 =  \left<T_0^{\ 0}\right>
 = -\left<T_r^{\ r}\right>$,
 and 
 the thick dotted curve represents
 $P_\scalar
 = -\left<T_\theta^{\ \theta}\right>
 = -\left<T_\varphi^{\ \varphi}\right>$.
 $\rho_\scalar$ is positive definite, but
 $P_\scalar$ becomes negative for $r \simeq r_\BH$.
 These curves depend only weakly on the parameters $q$ and $\alpha$.
 \label{Fig.FGrho.eps}%
 }%
\end{figure}

We obtained the solution for $q = 0.99999$ and $\alpha=1$.
The resultant radius of the solution is
\begin{eqnarray}
 r_\BH &=& 1.12871 \times l_\pl,
\end{eqnarray}
where $l_\pl := 1/m_\pl$ is the Planck length.
The frequency of the mode is
$\omega_{10} = 0.0253909 \times \alpha \: (1-q^2) \: m_\pl$.
The redshift factor $\alpha$ of the singularity and
the factor $(1-q^2)$, which is the shift of the Hawking temperature
\EQ{Hawking-temp.eq},
are directly reflected in the frequency $\omega_{10}$.
The form of the radial mode function for the scalar field is
displayed in \fig{Fig.f.eps},
where
the initial value of the radial mode function is
\begin{eqnarray}
 {f}_\sing &=& {f}(0) \ =\  1.268776 \times m_\pl.
\end{eqnarray}%
The distributions of the resultant metric elements $F(r)$ and $G(r)$
are displayed in \fig{Fig.FGrho.eps}-a.
To contrast our result with that for an ordinary charged black hole,
we also plot the exterior part $(r>r_\BH)$
of the Reissner-Nordstr\"om metric \EQ{RN-metric.eq}
with the same radius $r_\BH$ and the same charge $q$
(the thick dotted curves in \fig{Fig.FGrho.eps}-a).


\section{Many Particle Limit}

We consider a many quanta (particles) limit
in the structure~\cite{Nagatani:2003rj,Nagatani:2003new}.
In the many particle limit, 
it is natural to consider behavior of the thermal fluid
rather than that of the each motion of the fluctuation fields $\Phi$.
Thus we treat the quanta as a thermal fluid of the particles,
and this treatment simplifies the analysis.

The fluid is thermalized,
because the fluctuation should be time-independent and stationary.
We define local temperature of the thermal fluid as $T(r)$
due to spherical symmetry.
The thermal fluid is perfect one in the gravity background,
because 
we do not need to consider the interactions among the particles
except for the interaction with the mean field of the gravity.
Therefore the energy momentum tensor of the fluid is given by
$T_{\mu}^{\ \nu}(r) = \diag\left[\rho(r),-P(r),-P(r),-P(r) \right]$,
where $\rho(r)$ and $P(r)$ are energy density and pressure respectively.
The fluctuation fields are massless, then we have
\begin{eqnarray}
  \rho(r) &=& \frac{\pi^2}{30} g_* T(r)^4,
   \qquad
  P(r) = \frac{1}{3} \rho(r).
  \label{rho-P-def.eq}
\end{eqnarray}

Instead of solving the field equations for $\Phi$
in the space-time background $g_{\mu\nu}$,
we take {\em the proper temperature ansatz}, i.e.,
we assume the temperature distribution of the fluid
to be~\cite{Hotta:1997yj,Iizuka:2003ad,Nagatani:2003ps}
\begin{eqnarray}
 T(r) &=& \frac{T_\BG}{\sqrt{F(r)}},
  \label{proper-temp.eq}
\end{eqnarray}
where $F(r)$ is metric element $g_{00}$,
and $T_\BG$ is the background temperature.
This temperature distribution is natural
in the thermal equilibrium in the gravity background $g_{\mu\nu}$.
It is believed that
the temperature ansatz \EQ{proper-temp.eq} can be
approximately derived by using a combination of
the mean field approximation and 
the fluid approximation of the particles.

According to our concept of model construction,
the background temperature $T_\BG$
has been chosen to be the Hawking temperature $T_\BH$.
The background universe becomes Einstein static universe.

The Einstein equation (\ref{ein0.eq}) becomes following equations:
\begin{eqnarray}
  \frac{-G + G^2 + r G'}{r^2 G^2}
   &=&
   \frac{8 \pi}{m_\pl^2}
   \left\{\rho(r) + \rho_\BG + \frac{Q^2}{8\pi} \frac{1}{r^4} \right\},
   \label{Erho.eq}
   \\
  \frac{F - F G + r F'}{r^2 F G}
   &=&
   \frac{8 \pi}{m_\pl^2}
   \left\{P_r(r) - \rho_\BG - \frac{Q^2}{8\pi} \frac{1}{r^4} \right\},
   \label{EPr.eq}
\end{eqnarray}
We obtain the relation
\begin{eqnarray}
 G(r) &=&
  \left\{{1 - \frac{r}{2} \frac{\rho'(r)}{\rho(r)}}\right\}
  \left[
  {1 + \frac{8\pi}{m_\pl^2} r^2
   \left\{
    \frac{1}{3} \rho(r) - \rho_\BG - \frac{Q^2}{8\pi}\frac{1}{r^2}
   \right\} }
   \right]^{-1}
  \label{G-rho.eq}
\end{eqnarray}
from the equation \EQ{EPr.eq} with \EQ{rho-P-def.eq} and \EQ{proper-temp.eq}.
By substituting \EQ{G-rho.eq} into the equation \EQ{Erho.eq}
with \EQ{rho-P-def.eq},
we obtain the differential equation for the energy density $\rho(r)$ as
\begin{eqnarray}
 	&& 
		\:-\: 24 r 		\rho^2
		 \left\{ \textstyle
		  \rho  - \rho_\BG  + \frac{Q^2}{8\pi}\frac{1}{r^4}
		 \right\}
		\:+\: 12	r^2	\rho\rho'
		 \left\{ \textstyle
		  \rho + 2 \frac{Q^2}{8\pi}\frac{1}{r^4}
		 \right\}
	\nonumber\\
	&&
		\:+\:	r^3	\rho'^2
		 \left\{ \textstyle
		  \rho - 9 \rho_\BG - 9 \frac{Q^2}{8\pi}\frac{1}{r^4}
		 \right\}
		\:-\: 2 	r^3	\rho	\rho''
		 \left\{ \textstyle
		  \rho -3 \rho_\BG -3 \frac{Q^2}{8\pi}\frac{1}{r^4}
		 \right\}
	\nonumber\\
	&& \:+\: \frac{3 m_\pl^2}{8\pi}
	\left\{
		- 4 \rho \rho'
		+ 3 r \rho'^2
		- 2 r \rho \rho''
	\right\} \;=\; 0. \label{rhoEQ.eq}
\end{eqnarray}

The differential equation \EQ{rhoEQ.eq} is numerically solved
and we find out the solution
whose exterior part corresponds with the charged black hole
in the Einstein static universe.

\begin{figure}
 \begin{center}
  \begin{tabular}{@{}l@{}}
   \includegraphics[scale=0.45]{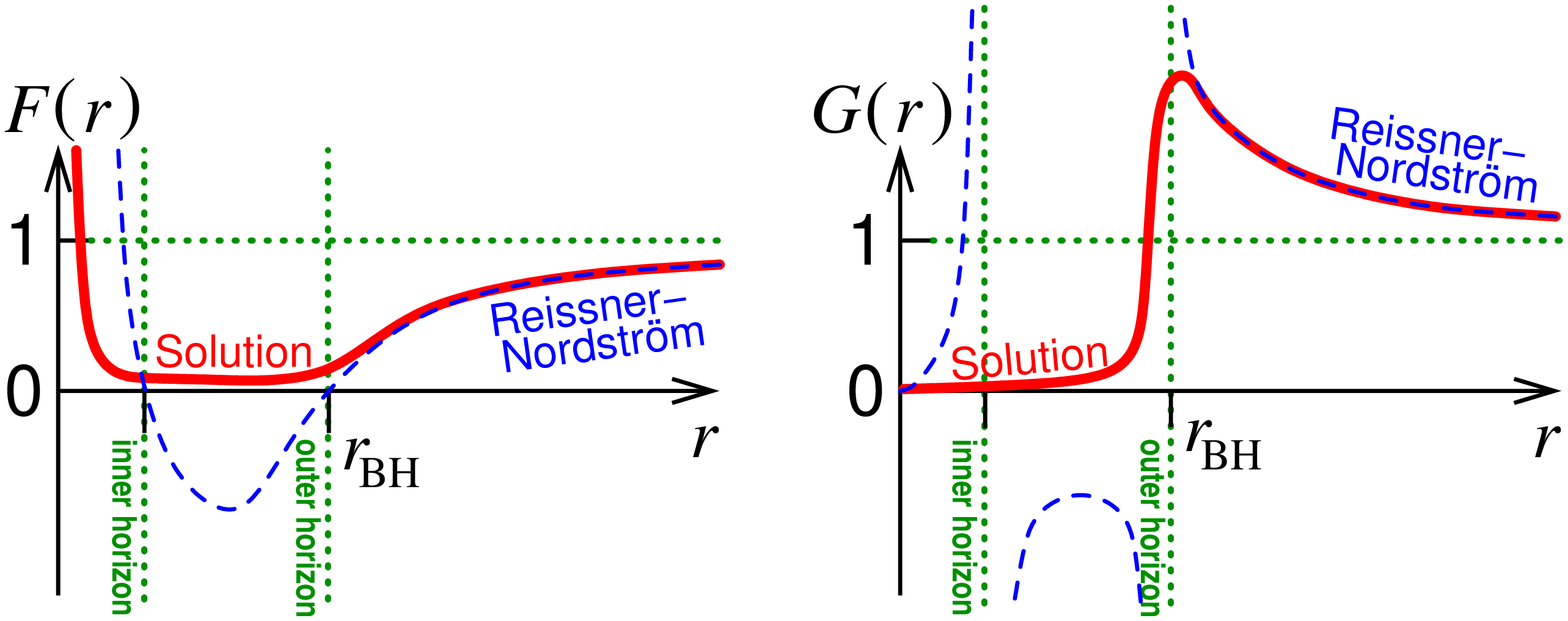}\\
   \includegraphics[scale=0.45]{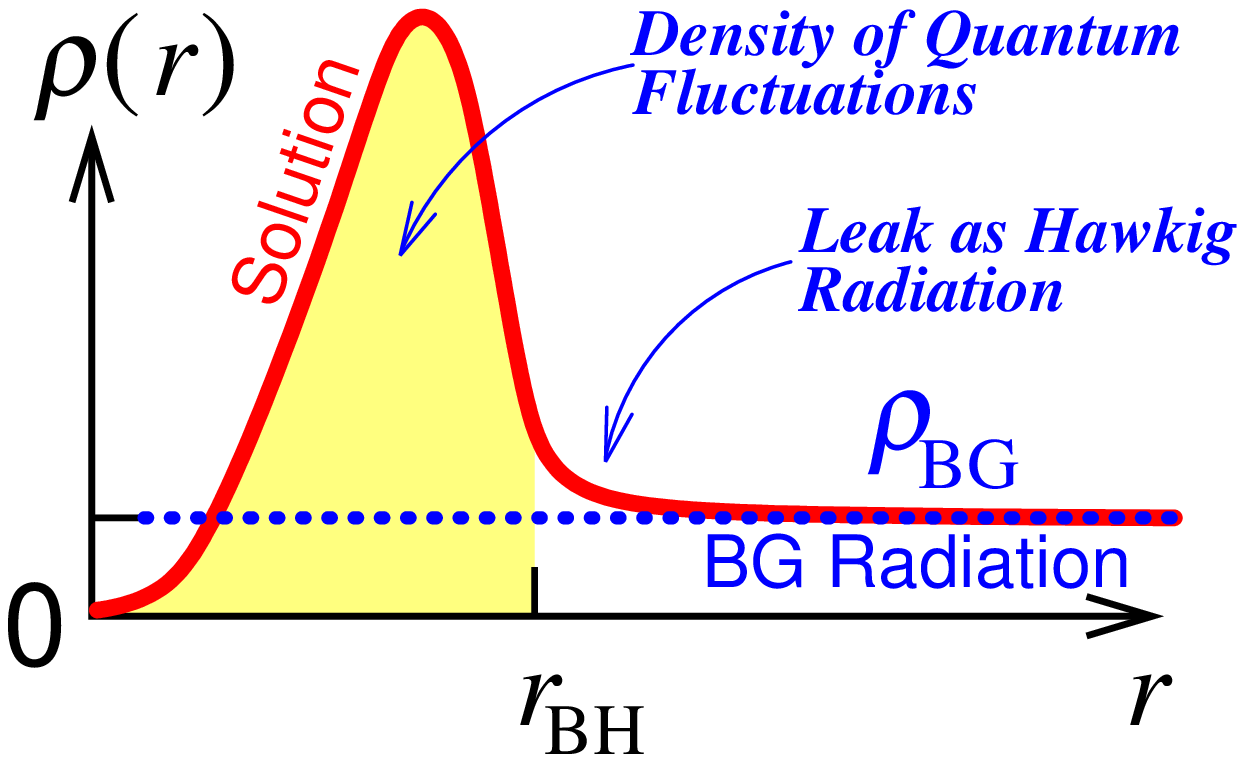}
  \end{tabular}
 \end{center}
 \caption{%
  Outline of the generic solution of the structure model with the many particle
 limit.
 The metric elements $F(r)$, $G(r)$ and the density distribution
 $\rho(r)$ are figured.
 The blue dotted curves indicate the Reissner-Nordstr\"om metric of
 the black hole corresponding to the model.
 \label{Fig.C-outline.eps}
 }%
\end{figure}

\begin{figure}
 \begin{tabular}{@{}c@{\hspace{8mm}}c@{}}
  \includegraphics[scale=0.4]{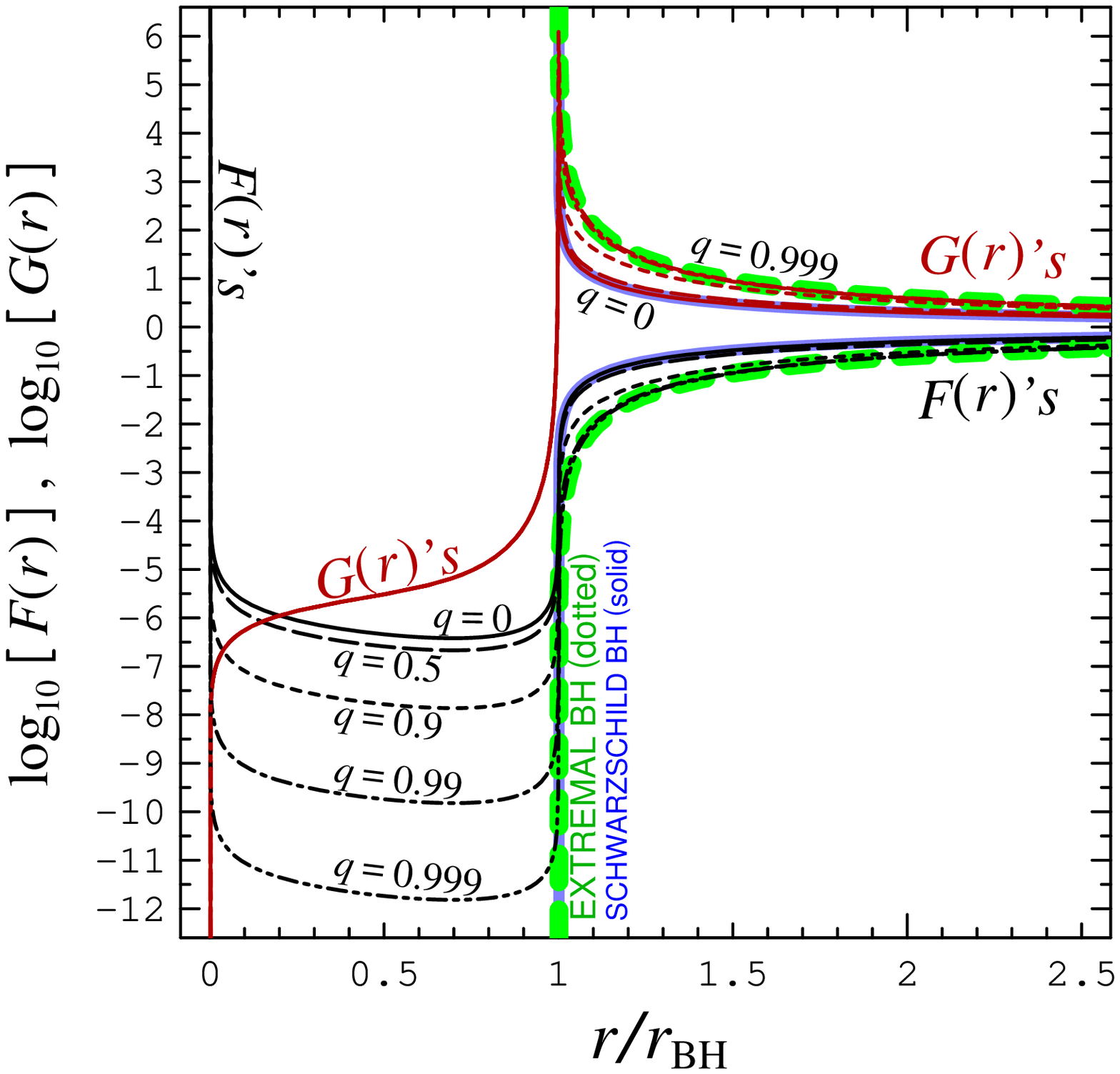} &
  \includegraphics[scale=0.4]{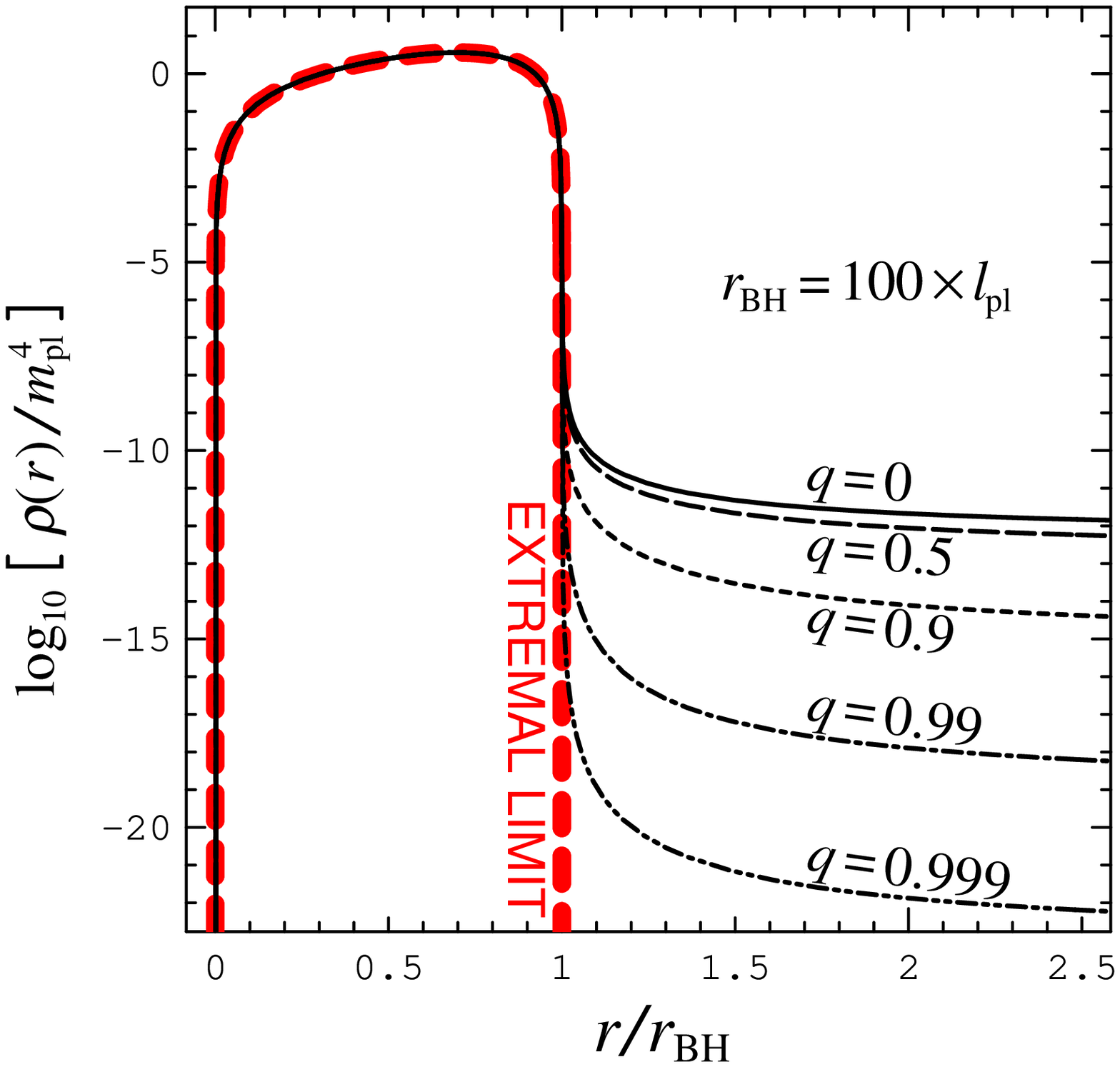} \\
  a) Elements of the metric $F(r)$ and $G(r)$ &
  b) Distributions of the energy density $\rho(r)$
 \end{tabular}
 \caption{%
 a) Elements of the metric $F(r)$ and $G(r)$
 and
 b) Distributions of the energy density $\rho(r)$
 of the quantum fluctuations,
 in the solution of the structure model with the many particle limit.
 The numerical solutions
 for the radius $r_\BH = 100\times l_\pl$, $g_* = 4$
 and for various charges $q = 0, 0.5, 0.9, 0.99$ and $0.999$
 are presented.
 The horizontal axis is the coordinate $r$
 normalized by $r_\BH$.
 The thick dotted (green) curves and the thick solid (blue) curves 
 indicate
 the exterior parts of 
 the extremal RN metric ($q \rightarrow 1$) and
 the Schwarzschild metric ($q = 0$) respectively.
 The distribution in the structure ($r < r_\BH$)
 has a universal form.
 The density $\rho(r)$ approaches to the background density $\rho_\BG$
 defined in \EQ{rho-BG.eq} when $r$ becomes large.
 We find that the function form of $G(r)$ in the structure
 is almost independent of $q$.
 \label{CFGrho.eps}%
 }%
\end{figure}

The solution is parameterized by $r_\BH$ and $q$
which are the outer horizon radius and the charge-parameter
of the correspondent RN black hole respectively.
The element of the metric $F(r)$ is derived by
the proper temperature ansatz \EQ{proper-temp.eq}
and $G(r)$ is derived by \EQ{G-rho.eq}.
The outline of the solution is shown in \fig{Fig.C-outline.eps},
and typical numerical solutions
are displayed in \fig{CFGrho.eps}.
The solution indicates that
most of the quantum fluctuations are trapped in the structure $(r < r_\BH)$
by the gravitational potential $F(r)$,
and that the radius of the structure corresponds with that of
the correspondent black hole, namely, $r_\BH$.
The inner horizon of the RN solution becomes completely meaningless.


\section{Information of Black Hole and Decay of Baryon Number}

In our structure model of black holes,
any information of the structure is carried by the quantum fluctuation
around the singularity.
Therefore the Bekenstein entropy
is explained by counting of the states
of the quantum fluctuation~\cite{Nagatani:2003rj,Nagatani:2003new}.

We have obtained the temperature distribution $T(r)$ of the thermal fluid
in the many body limit,
thus the entropy density distribution can be calculated
as $s = \frac{2\pi^2}{45} g_* T^3$.
We calculate the total entropy of the structure as
\begin{eqnarray}
 S
  &\simeq&
  \int_{0}^{r_\BH} 4 \pi r^2 \sqrt{G(r)} \;
  s\left(T(r)\right)
 \;=\;
  \frac{(8 \pi)^{3/4}}{\sqrt{5}} m_\pl^2 r_\BH^2
 \;\simeq\; 5.0199 \times \frac{r_\BH^2}{l_\pl^2}.
 \label{entropy-result.eq}
\end{eqnarray}
This result reproduces the area law of the Bekenstein entropy,
and is a little greater than the the Bekenstein entropy:
$S/S_{\rm Bekenstein} \simeq 1.5978$.
This slight difference seems to be caused by the approximation
of the many body limit.

The baryon number is also carried by the internal quantum fluctuation
in the same way of the entropy~\cite{Nagatani:2003new2}.
Therefore we can define the baryon number density $b(r)$
in the structure,
and we calculate the total baryon number of the structure as
\begin{eqnarray}
 B &:=&
  \int_{0}^{r_\BH} 4 \pi r^2 dr \sqrt{G(r)} \; b(r)
 \label{B.eq}
\end{eqnarray}
which is regarded as {\it the baryon number of the black hole}.

The Higgs scalar vev $\left<\phi\right>$ in the structure vanishes
and the symmetry of the electroweak (EW) theory restores
because the temperature in the structure ($\sim$ Planck scale)
is much greater than the EW scale ($\sim100\GeV$)
and the thermal phase transition of the EW theory arises
\cite{Nagatani:1998gv,Nagatani:2003pr,Nagatani:2001nz}.
Therefore the sphaleron processes are working in the structure,
and are changing the baryon number.
The sphaleron transition rate for the proper-time is given by
$\Gamma_\sph = \kappa \alpha_\weak^4 T^4$,
where $\kappa\sim O(1)$ is a numerical constant and
$\alpha_\weak$ is the weak gauge coupling constant
at the Planck energy
\cite{Arnold:1996dy:Huet:1996sh:Moore:1997sn}.
When we assume GUT, the GUT interaction also violates the baryon number
as the rate $\Gamma_\GUT = \kappa \alpha_\GUT^2 T^4$.
The time-evolution of the baryon number density
on the coordinate-time $t$ is given
by the Boltzmann-like equation \cite{Garcia-Bellido:1999sv}:
$ \frac{d}{dt}b = - \frac{39}{2} \sqrt{F} \frac{\Gamma}{T^3} b $.
By using the proper temperature ansatz \EQ{proper-temp.eq},
the Boltzmann-like equation becomes
\begin{eqnarray}
 \frac{d}{dt}b &=& - {\cal R} b
 \label{Boltzmann2.eq}
\end{eqnarray}
where we have defined the baryon number decay rate:
\begin{eqnarray}
 {\cal R} &=&
  \left\{
   \begin{array}{lll}
    \frac{39}{2} \kappa \alpha_\weak^4 T_\BH &
     \simeq 10^{-6} / r_\BH & \mbox{(Sphaleron)} \\[0.2em]
    \frac{39}{2} \kappa \alpha_\GUT^2 T_\BH &
     \simeq 10^{-3} / r_\BH & \mbox{(GUT)}
   \end{array}
  \right..
  \label{b-decay-curve.eq}
\end{eqnarray}
For an astrophysical black hole $(r_\BH \sim 1 {\rm km})$,
the baryon number decay time becomes
\begin{eqnarray}
 1/{\cal R} &=&
  \left\{
   \begin{array}{ll}
    \mbox{a few seconds}      & \mbox{(Sphaleron)} \\
    \mbox{a few mili-seconds} & \mbox{(GUT)}
   \end{array}
  \right.,
\end{eqnarray}
therefore, we find a rapid decay of the baryon number
in the black hole~\cite{Nagatani:2003new2}.

\section{Conclusions}

Our picture of a black hole is quite similar to
that of an atom
consisting of a charged nucleus and a quantized electron field
of several quanta.
The radius of the ball structure, namely the Schwarzschild radius,
is quantized
due to the quantization of the fluctuating field in the structure.
Particularly we construct the minimum structure of the black hole.
The field fluctuation in the minimal model
contains only one quantum in the 1s-wave mode.
The minimum model is quite similar to that of a hydrogen atom,
whose electron is represented by the 1s-wave function.
The radius of the minimum model,
namely the minimum Schwarzschild radius, is approximately $1.1287$
times the Planck length.
There is an obvious analogy between the minimum Schwarzschild radius
in our model and the Bohr radius of the hydrogen atom.

At the present time,
a quantization of black holes seems to be difficult,
because we have no correct theory of quantum gravity.
However,
the Bohr's model for atomic structure in history of old quantum theory
is instructive:
Bohr succeeded in quantizing the hydrogen atom
by his structure model of the atom
before proposition of the Schr\"odinger equation.
Without a theory of quantum gravity,
we propose the quantization of black holes
by the structure model of black holes based on the Einstein gravity.
Our aim is just to construct {\em the Bohr's quantization} of black holes.

\begin{figure}
 \begin{center}
  \includegraphics[width=8cm]{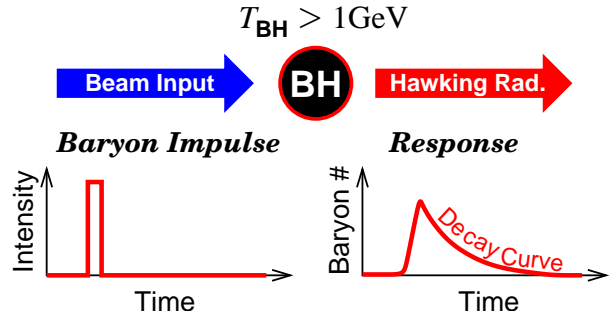}%
 \end{center}
 \caption{%
 The experimental verification of the structure model.
 \label{Fig.BPulse.eps}%
 }%
\end{figure}

Our model can be verified by the experiment
which is schematically shown in \fig{Fig.BPulse.eps}.
The Hawking radiation contains the internal baryon number,
because the Hawking radiation is regard as a leak
of the internal quantum fluctuations.
If we observe the baryon decay curve (\ref{b-decay-curve.eq})
by the following experiment:
1) prepare a small black hole ($T_\BH > 1\GeV$)
which can radiate protons,
2) input a baryon beam pulse into the black hole, and
3) observe the baryon number in the Hawking radiation,
then we confirm our structure model of the black hole.

\section*{Acknowledgements}

 I would like to thank
 Ofer~Aharony, Micha~Berkooz, Hiroshi~Ezawa, Satoshi~Iso, Hiraku~Kawai,
 Barak~Kol, Masao~Ninomiya, Kei~Shigetomi and Kunio~Yasue
 for useful discussions.

%

\end{document}